# Ice thickness measurements by Raman scattering

Sergey M. Pershin,[1] Vasily N. Lednev,[1] Vladimir K. Klinkov,[1] Renat N. Yulmetov,[1,2] and 1Alexey F. Bunkin[1]

[1]*Wave Research Center, Prokhorov General Physics Institute, Russian Academy of Science, Vavilov str., 38, Moscow, Russia*
[2]*The University Centre in Svalbard, Longyearbyen, Norway*
*Corresponding author: pershin@kapella.gpi.ru*

**Abstract:** A compact Raman LIDAR system with a spectrograph was used for express ice thickness measurements. The difference between the Raman spectra of ice and liquid water is employed to locate the ice-water interface while elastic scattering was used for air-ice surface detection. This approach yields an error of only 2 mm for an 80-mm-thick ice sample, indicating that it is promising express noncontact thickness measurements technique in field experiments.

There is growing scientific and technological interest in the Polar Regions due to Global climate change and increased industrial activity in the Northern areas. The Arctic Ocean is an excellent indicator of the climate change and at the same time has a strong global impact. In the Polar Regions ice serves as an effective interface between the ocean and the atmosphere, restricting heat transfer, mass, momentum and chemical constituents. In light of this, ice studies play an important role in global climate modelling and forecasting. However, reliable prediction of future climate changes can be made only if consistent data on different physical parameters of ice and water are available. These parameters include ice thickness, temperature, salinity, surface roughness, optical and thermodynamic properties of ice and snow. Conventional techniques for ice thickness, temperature and salinity measurements are currently time and manpower consuming [1] and every single point measurement needs several hours to perform. Satellite measurements of ice thickness have suffered from low accuracy (20% for ice thickness below 20 cm and 100% for 50 cm layer) and are strongly influenced by temperature and salinity of seawater [2,3]. A good way to solve this problem is to develop a compact LIDAR (Light Detection and Ranging) device [4,5] installed on unmanned aircraft or autonomous underwater vehicle which can perform fully automatic measurements of ice properties in the Polar areas.

Another motivation for express ice thickness measurements in the Arctic region is hydrocarbons exploration and exploitation. Ice loads estimation and the probability of their occurrence are key points in the Arctic offshore objects design. Ice loads depend on thermo-mechanical properties of sea ice such as thickness, temperature, salinity [6]. Monitoring of ice conditions in the region of interest provides valuable information about variation of the properties and therefore about possible loads on port constructions.

There is a vast variety of ice thickness measurement methods including drilling, electromagnetic sounding (airborne or ground-based), Differential GPS surveying with total-station, mass-balance buoys, various sonars, and laser altimetry (ground based, airborne, satellite) [1]. The methods differ in terms of accuracy, spatial and time resolution, coverage, and cost. Drilling is the most popular method due to its relative simplicity and high accuracy (centimetres). However, drilling is a single-point measurement method, meaning it is time consuming and requires work on the sea ice. Electro-magnetic sounding gives much higher sampling frequency but lower accuracy, which drops dramatically for the sea ice with high porosity. Upward looking sonars are costly and usually deployed for long term measurements. The remaining measurement methods are less popular or yield insufficient information about ice.

In this paper we suggested an alternative technique for ice thickness measurement, using optical methods previously used for remote sensing of the ocean [7,8]. A laser rangefinder can easily detect the top border of ice through elastic scattering while the ice bottom interface (ice-water) is a nutshell due to similar refraction indexes for ice and water. The Raman spectra of ice and water differ substantially in terms of their OH-band profile and this feature can be used to detect ice-water interface. As previously shown, the Raman scattering can be used for thin films thickness measurements [9-,10,11] while in laser remote sensing Raman spectroscopy is a promising technique for the detection of different parameters of water or ice (temperature, salinity etc.) [12-,13,14,15,16,17,18,19]. We suggested a new simple and convenient non-contact express optical method for ice thickness measurements which extends the applications of Raman spectroscopy.

The experimental setup is presented in Fig. 1. The compact Raman LIDAR system was described in detail in our previous paper [8] and here will be described shortly. The system is based on a compact diode pumped solid state YVO4:Nd laser (527 nm, 5 ns, 1 kHz, 200 J/pulse, beam quality M2 = 1.2; beam diameter at 1/e2 level 1.8 mm). Laser irradiation scattered by remote objects are collected by a quartz lens (F = 21 cm) to the input slit of the spectrograph. The detection system consists of compact spectrograph (Spectra Physics, MS127i) equipped with gated detector (ICCD, Andor iStar). The chosen spectral window was 500 - 750 nm, thus making it possible to register several signals simultaneously: elastic scattering (527 nm) (Mie and Rayleigh scattering) and Raman scattering (OH-band centre 640 nm). The signals for elastic and Raman scattering were determined as the integral of the corresponding band with background correction. The spectra were summed by 100 laser pulses to improve reproducibility of measurements. The ICCD detector allowed us to obtain gated images with 5 ns duration. The time jitter between laser pulse and detection gate was less than 3 ns. The effective depth resolution was estimated as 1.2 m in air. As such, the time-of-flight technique could not be used for the thickness measurements of samples with depth lower than several metres. We used an alternative approach for ice thickness measurement with lens-to-sample distance variation and the detection of signals from different laser beam waist positions. The choice of optimal focal length for the lens used in experiments was formulated by two conditions: lens focal length should be greater than sample thickness to measure both ice boards (air-ice and ice-water) and lens focal length should be as small as possible to minimise beam waist and improve spatial resolution. For this purpose we used a quartz lens (F = 210 mm, diameter 5 cm) mounted on a movable stage. The laser beam waist was 65 μm in diameter and estimated Rayleigh length was 0.55 mm.

The distilled water was frozen down to -20 $^0$C in a laboratory refrigerator for 24 hours to form an ice brick which was of good optical quality with a few defects inside (bubbles, brines, cracks etc.). An ice sample was then placed to float on the top of a tank filled with distilled water at +0.5 $^o$C.

The typical Raman spectra of water and floating ice surfaces at temperature near 0 $^0$C are presented in the Fig.2. The spectra were normalised on intensity of elastic scattering line for better view. The intensity of elastic scattering increased dramatically for the ice sample compared to liquid water due to defects and air bubbles at the upper ice surface. The detailed Raman OH-band profiles for ice and water are presented on inset in the figure. The ice spectrum was obtained for beam waist located in the centre of the ice sample. The water spectrum was detected through the floating ice layer for beam waist located at distance of 30 mm below ice- water interface. The detailed spectra for the Raman OH-band profiles are typical for ice and water and this feature can be used to detect ice-water boundaries. A Raman spectrum for the laser beam waist located near ice-water interface will represent a sum of water and ice signals, thus meaning that a procedure for the quantitative detection of OH-band profile is needed. The conventional way of detecting OH-band profile distortion is to use nonlinear fitting by two or multiple peaks and then take a ratio of the integral peaks as a measure of profile variation. Such a procedure has been already used for remote measurements of water temperature by means of a Raman OH-band profile [15,20].

As previously shown, small distortion of the OH-band profile can be easily detected by curve fitting with symmetrical function (Gaussian or Lorentz) and comparison of curve's centres [19]. Our approach resulted in improved accuracy of remote temperature measurements by means of a Raman OH-band profile [13,19]. In this study the same procedure (curve centres comparison) was used to quantify OH-band profile distortions. The ice and water Raman OH-band profiles were fitted with Gaussian function (thin curves in Fig. 2) and results are shown inset: corresponding centres differed substantially (3206 vs 3271 cm-1). Consequently, the centre of the OH-band profile can stand as a good quantitative indicator for the detection of ice-water interface.

We have detected a series of Raman spectra for different positions of laser beam waist by varying the lens-to-sample distance. The refractive index of ice (n = 1.330) will result in continuous increase in lens to beam waist position when the lens is moving towards the ice surface. The correction on longer beam waist position was made and the corresponding axis is presented in Fig.3 (black colour). The Raman and elastic scattering signals (integrals) and centre of OH-band profiles were calculated from the spectrum at every lens position and the results are presented in Fig. 3. The intensity of elastic scattering (Fig. 3 a) can be successfully used for the detection of air-ice interface, although it fails to detect ice-water boundaries. Both substances have the same refractive indexes (1.330 and 1.333 correspondingly for ice and water) and high optical quality of ice-water interface (air bubbles or defects free due to slow growth of ice in water at equilibrium condition). Indeed, this resulted in the absence of elastic scattering.

The Raman OH-band profile was fitted with a Gaussian function and corresponding curve centres were plotted as a function of lens-to-sample distance. The s-type dependence of centres on lens-to-sample distance clearly showed an ice-to-water boundary. The ice thickness was estimated as the distance between maximum elastic scattering signal and half maximum change of OH-bond profile centre's dependence (dashed vertical lines in Fig.3). The accuracy of the ice thickness measurements is determined by sum errors for air-ice and ice-water interface detection. The detection accuracy for air-ice border was calculated with the FWHM of the registered elastic signal (Fig. 3a). The accuracy of ice-water interface was estimated by OH-band profile centres as follows: a group of data near inflection point was fitted with linear regression and accuracy was stated as an interval between confidence band interval at inflection point (Fig. 3b). This resulted in accuracy of ±2 mm. The measured ice thickness was 81 mm, which fit fairly well with the 80 mm value measured by the slide-meter.

This study has suggested a simple express procedure for optical non-contact ice thickness measurements by means of Raman spectroscopy. The Raman OH-band spectra difference for water and ice samples was used as a key indicator of ice-water border while elastic scattering was used for air-water interface detection. The estimated accuracy of the measurements was 2 mm for 80 mm thick ice floating in water. However moving lens system is time consuming and laborious and for field measurements the compact device without movable parts are preferred. The technique for simultaneous measurements of Raman spectra at multiple distances [21] can be used to solve this problem and to develop a compact lidar system (<1 kg) for ice thickness measurements.

*This study has been partially supported by the Russian Foundation for Basic Research (RFBR) projects 14-02-00748, 14-02-00018, the Russian President Program for Leading Scientific Schools 4482.2014.2 and the Russian Academy of Sciences Program №28*

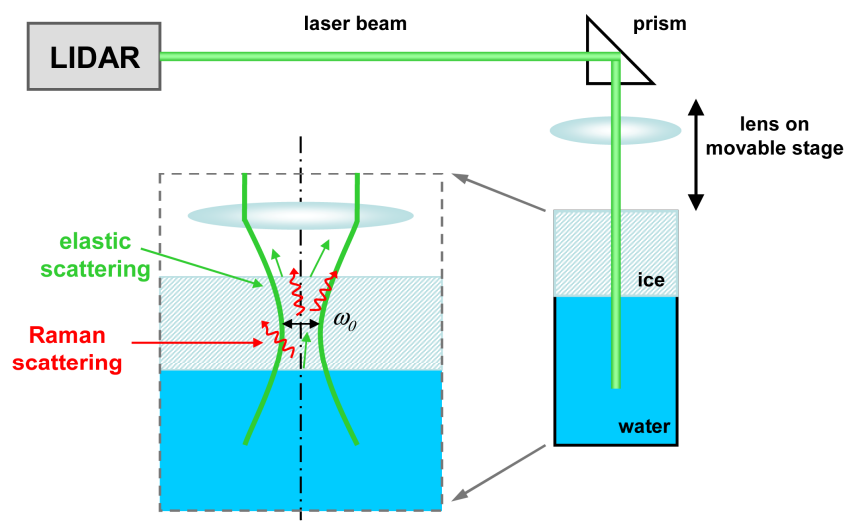

**Fig. 1** Experiment setup for ice thickness measurements using Raman OH-band profile. (The movable lens is used to adjust the laser beam waist position in the ice-liquid water sample.)

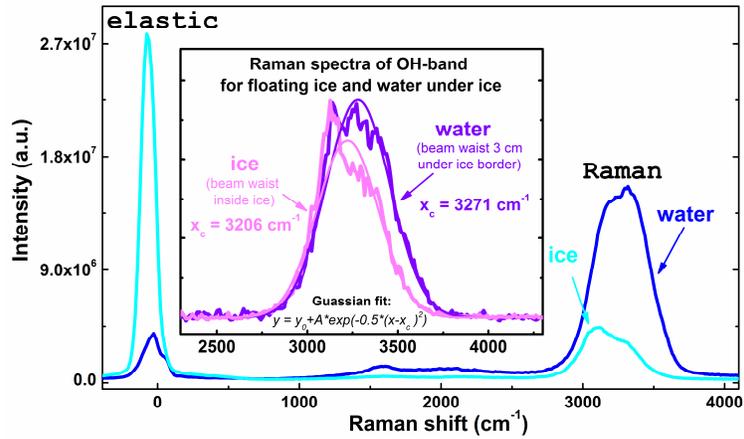

**Fig. 2** The ICCD registered spectra: cyan, laser beam waist adjusted to the ice sample center; blue, laser beam waist adjusted 30 mm below the ice-liquid water interface. The strong elastic signal is used for air-ice interface identification. The inset plot gives detailed Raman water spectra in the OH stretching range. The fitted Raman spectra centers have a difference (3206 cm$^{-1}$ for ice and 3271 cm$^{-1}$ for liquid water) which can be used to determine ice-liquid water interface.

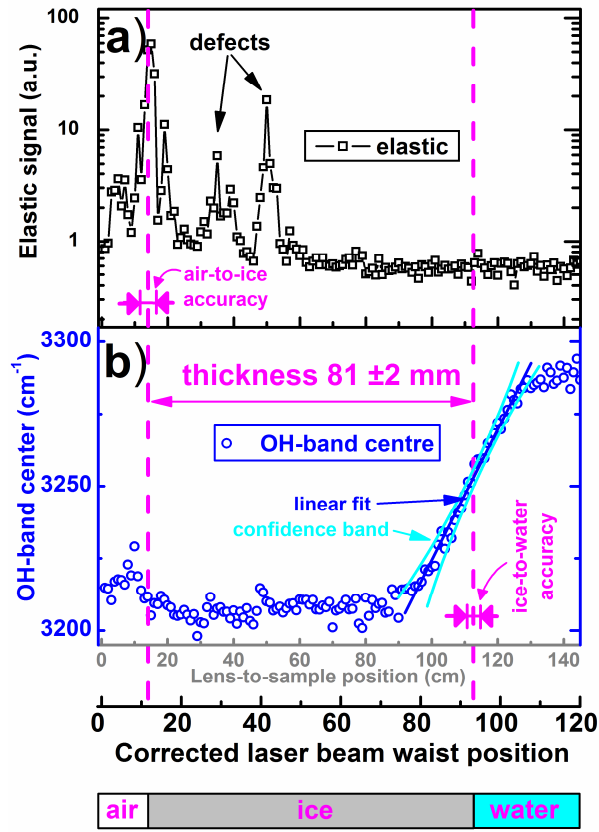

**Fig.3** The ice thickness measurements using elastic and Raman scattering:
a) Ice thickness measurement using elastic signal (black squares);
b) Ice thickness measurements using Raman OH-band centre (blue circus).
The lens was shifted to the floating ice sample (x axis – grey colour) and the spectrum was detected at every position. The beam waist position change due to refractive index and optical path in the ice sample was corrected (x axis – black colour) in order to determine ice sample thickness.
The elastic signal was determined as integral to the corresponding band with linear background correction. The OH-band profile was fitted with Gaussian function and the curve centres as a function of lens-to-sample distance were plotted in Fig.3 b.